\documentclass[12pt,preprint]{aastex}
 
\usepackage{graphicx}
\usepackage{epstopdf}
\usepackage{psfig}

\slugcomment{Submitted to the ApJ Letters EVLA special issue}
 
\shorttitle{Molecular proto-cluster}
\shortauthors{Carilli et al.}
 
\begin{document}
  
 \title{EVLA observations of a proto-cluster of
molecular gas rich galaxies at $z = 4.05$}
 
\author{ 
C.L. Carilli\altaffilmark{1},
J. Hodge\altaffilmark{2},
F. Walter\altaffilmark{2},
D. Riechers\altaffilmark{3},
E. Daddi\altaffilmark{4},
H. Dannerbauer\altaffilmark{2},
G.E. Morrison\altaffilmark{5}
}

\altaffiltext{$\star$}{The Very Large Array of the National Radio Astronomy
Observatory, is a facility of the National Science Foundation
operated under cooperative agreement by Associated Universities, Inc}

\altaffiltext{1}{National Radio Astronomy Observatory, P.O. Box 0, 
Socorro, NM, USA 87801-0387}
\altaffiltext{2}{Max-Planck Institute for Astronomy, Konigstuhl 17, 69117,
Heidelberg, Germany}
\altaffiltext{3}{Department of Astronomy, California Institute of Technology,
MC 249-17, 1200 East California Boulevard, Pasadena, CA 91125, USA; Hubble 
Fellow}
\altaffiltext{4}{Laboratoire AIM, CEA/DSM - CNRS - University Paris Diderot, 
DAPNIA/Service Astrophysccique, CEA Saclay, Orme des Merisiers, 
91191 Gif-sur-Yvette, France}
\altaffiltext{5}{Institute for Astronomy, University of Hawaii, Honolulu, HI,
96822, USA ; Canada-France-Hawaii Telescope, Kamuela, HI 96743}

\begin{abstract}

We present observations of the molecular gas in the GN20 proto-cluster
of galaxies at $z =4.05$ using the Expanded Very Large Array (EVLA). This
group of galaxies is the ideal laboratory for studying the formation
of massive galaxies via luminous, gas-rich starbursts within 1.6 Gyr
of the Big Bang.  We detect three galaxies in the proto-cluster in CO
2-1 emission, with gas masses (H$_2$) between $10^{10}$ and $10^{11}
\times (\alpha/0.8)$ M$_\odot$. The emission from the brightest
source, GN20, is resolved with a size $\sim 2"$, and has a clear
north-south velocity gradient, possibly indicating ordered rotation.
The gas mass in GN20 is comparable to the stellar mass ($1.3\times
10^{11} \times (\alpha/0.8)$ M$_\odot$ and $2.3\times 10^{11}$
M$_\odot$, respectively), and the sum of gas plus stellar mass is
comparable to the dynamical mass of the system ($\sim 3.4\times
10^{11} [sin(i)/sin(45^o)]^{-2}$ M$_\odot$), within a 5kpc radius.
There is also evidence for a tidal tail extending another $2"$ north
of the galaxy with a narrow velocity dispersion.  GN20 may be a
massive, gas rich disk that is gravitationally disturbed, but not
completely disrupted. There is one Lyman-break galaxy (BD29079) in the
GN20 proto-cluster with an optical spectroscopic redshift within our
search volume, and we set a 3$\sigma$ limit to the molecular gas mass
of this galaxy of $1.1\times 10^{10} \times (\alpha/0.8)$ M$_\odot$.

\end{abstract}
 

\section{Introduction}

Numerous lines of evidence support the hypothesis that massive
elliptical galaxies form the majority of their stars quickly at early
epochs, including: stellar population synthesis studies of nearby
elliptical galaxies (Renzini 2006; Collins et al. 2009), a faster
decline with cosmic time in the star formation rate per unit stellar
mass with increasing galaxy mass (Moresco et al. 2010), and the direct
observation of `red and dead' ellipticals in clusters at high redshift
(Kurk et al. 2009; Doherty et al. 2009; Andreon \& Huertas-Company
2011).  Plausible progenitors for these evolved galaxies at high
redshift are even higher redshift submm galaxies (SMGs), corresponding
to dusty, luminous starburst galaxies found in wide-field submm
surveys (Blain et al. 2002). These hyper-luminous high-z galaxies may
trace high over-densities, and are likely related to the formation of
clusters and large ellipticals (Stevens et al.  2003; Aravena et
al. 2010; although cf. Chapman et al.  2009; Williams et
al. 2011). While the space density of SMGs peaks around $z \sim 2.3$
(Chapman et al. 2003), it is becoming clear that there is a
substantial high-redshift tail of the SMG population, extending to $z
> 4$ (Capak et al. 2011, Riechers et al. 2010; Wardlow et al. 2011).
Daddi et al. (2009a) conclude, based on SMG space densities and duty
cycles, that there are likely enough SMGs at $z > 3.5$ to account for
the known populations of old massive galaxies at $z \sim 2$ to 3.

A key question for the SMGs is: what drives the prolific star
formation?  Tacconi et al. (2006; 2008) argue, based on imaging of
higher-order CO emission from a sample of $z \sim 2$ SMGs, that SMGs
are predominantly nuclear starbursts, with median sizes $< 0.5"$ ($<
4$kpc), `representing extreme, short-lived, maximum star forming
events in highly dissipative mergers of gas rich galaxies.'  This
conclusion is supported by VLBI imaging of the star forming regions in
two SMGs (Momjian et al. 2005; 2010). However, recent imaging of the
lower order CO emission in a few SMGs (Ivison et al. 2010; Carilli et
al. 2010; Riechers et al. 2011 in press), suggests that the
lower-excitation molecular gas reservoirs can be significantly more
extended. 

Daddi et al. (2009a) have identified a unique region in GOODS-N,
containing the three galaxies: GN20, GN20.2a, and GN20.2b, at $z \sim
4.05 \pm 0.01$, within 20$''$ of each other (projected physical
separation = 140 kpc).  GN20 is among the brightest submm galaxies
known, with a 350 GHz flux density of 23mJy. The galaxies GN20.2a and
b are separated by only a few arcseconds, and hence are not spatially
distinct in typical submm bolometer images. The two galaxies have a
summed flux density of $S_{350\rm GHz} \sim 9.9$mJy. The implied IR
luminosities (8 to 1000$\mu$m) are $L_{TIR} \ge 10^{13}$
L$_\odot$, with star formation rates are $> 1000$ M$_\odot$
yr$^{-1}$ (Daddi et al. 2009a). High order CO emission (4-3 and/or
6-5) has been detected from all three galaxies, although only
marginally in the case of GN20.2b (Daddi et al. 2009a).
 
The GN20 field also contains numerous Lyman Break Galaxies (LBGs) at
similar redshift, and an overdensity of $z_{phot} > 3.5$ IRAC selected
galaxies (Daddi et al. 2009a). Hence, this field offers an
unprecedented opportunity to perform a definitive study of the gas
distribution, kinematics, and physical conditions in a proto-cluster
of galaxies within 1.6 Gyr of the Big Bang. In this paper we present a
survey with the Expanded Very Large Array (EVLA; Perley et al. 2011),
of the CO 2-1 emission from the GN20 proto-cluster.  The low order
transitions are critical for determining the total gas mass. We also
search for CO emission from other galaxies in the cosmic volume
surveyed.

\section{The GN20 proto-cluster: a massive galaxy formation
laboratory}

The HST I-band image of the bright SMG GN20 shows diffuse emission
about 1.5$"$ in extent (Daddi et al. 2009a), although offset from the
radio and submm emission by $\sim 1"$, implying high obscuration of
the most active star forming regions in the galaxy.  GN20 is detected
at 1.4 GHz, and resolved on a scale of $\sim 1.5"$, with a total flux
density at $72\pm 13\mu$Jy (Morrison et al. 2010; Casey et al. 2009).
High resolution imaging of the 850$\mu$m emission shows resolved
structure, with a north-south extension possibly as large at 1.5$"$
(Younger et al.  2008; Iono et al. 2006). The 6.2$\mu$m PAH spectral
feature has been detected in GN20 using Spitzer (Riechers et al. in
prep). The integrated CO 1-0 and 2-1 emission from GN20 has been
imaged at resolutions down to $0.2"$ with the VLA (Carilli et
al. 2010). The molecular gas is extended on a scale of at least
1.5$"$, and at high resolution forms a partial ring, or
disk. Unfortunately, the old VLA correlator provided no velocity
information. The excitation of the integrated CO emission from GN20 is
higher than the Milky Way, but lower than high redshift quasar hosts
and the nuclear starburst regions of IR-luminous galaxies. The CO 4-3
line strength in GN20 is more than a factor of two lower than expected
for thermal excitation (Carilli et al. 2010).

GN20.2a shows a complex and extended morphology in the HST images
(Daddi et al.  2009a), and it is a relatively strong radio source
($S_{1.4} = 181\mu$Jy), likely corresponding to a radio AGN. The SED
in the optical through near-IR is consistent with a star forming
galaxy.  GN20.2a has been detected in CO 4-3 emission by Daddi et
al. (2009a).  Daddi et al. derive an IR luminosity (8 to 1000$\mu$m) of
$L_{IR} = 1.6\times 10^{13}$ L$_\odot$.

GN20.2b is faint and compact in the HST image. It is also detected in
the radio continuum at $S_{1.4} = 32\mu$Jy, consistent with star
formation. CO 4-3 emission is marginally detected with the Plateau de
Bure Interferometer, and appears to be broad (700 km s$^{-1}$;
Daddi et al. 2009a).

Daddi et al. (2009a) find 15 B-band dropout galaxies in a 25$''$
radius centered on GN20, an overdensity of a factor of 6 compared to
the full GOODS-N area, which is significant at the 7$\sigma$ level.  A
spike in the redshift distribution of galaxies at $z=4.06\pm0.02$ is
observed in all of GOODS-N (13 spectroscopic redshifts in total at
this redshift).  Lastly, the SMG GN10 at $z = 4.04$ is located 9$'$
from GN20 (Daddi et al. 2009b; Dannerbauer et al. 2008).  Therefore,
it appears that the GN20 volume has a very significant overdensity,
indicating a proto-cluster environment at $z\sim 4.05$. Daddi et
al. (2009a) estimate a total mass for this structure of $\sim 10^{14}$
M$_\odot$.
 
\section{Observations} 

We observed the GN20 field with the Expanded Very Large Array in the D
(1km) configuration in March and April, 2010. Observations were made
of the CO 2-1 line, using a total bandwidth of 246 MHz and in each of
two polarizations, centered at 45.655 GHz.  The velocity coverage is
1600 km s$^{-1}$, including the CO 2-1 lines in the three
galaxies. The total observing time was 28 hours, but only about 22 to
24 antennas were available at a time due to on-going EVLA work.

Dynamic scheduling to ensure good weather is now standard at the EVLA.
Fast switching phase calibration was employed (Carilli \& Holdaway
1999) on timescales between two and three minutes using the VLA
calibrator J1302+5748.  Data were edited to remove time ranges of poor
phase stability.  The source 3C286 was used for flux density
calibration. Standard EVLA data calibration and editing was performed
using AIPS.  After calibration and data editing, images were
synthesized using the robust weighting scheme of the uv data with R=2
(Cornwell et al. 1999).

The final resolution for the images was $1.9"$. Spectral line cubes
were generated at 12MHz per channel (78 km s$^{-1}$).  The rms noise
per channel at 12MHz resolution is 0.11 mJy beam$^{-1}$. All images
were corrected for the VLA primary beam response, which has a FWHM
$\sim 1'$ at 45 GHz. The field pointing center was located
$10"$ west of GN20.

\section{Results} 

Figure 1 shows the CO 2-1 emission integrated over the full frequency
range covered by these observations for the GN20 group. The
crosses mark the 1.4 GHz continuum positions of the three galaxies
GN20, 20.2a,b (Morrison et al. 2010), and the optical position of one
LBG with a spectoscopic redshift in the redshift range of the CO 2-1
observations (Section 5.2). We have added the previously published D
array observations (Carilli et al. 2010) to improve signal-to-noise.
Unfortunately, the previous data did not have the exact same velocity
coverage, and hence the velocity-integrated flux densities are not
accurate. Hence, this image acts as a finding chart for the CO
emitting regions, but accurate line fluxes will be based on subsequent
analysis of the new EVLA data itself.

We detect CO 2-1 emission from GN20, GN20.2a, and GN20.2b. GN20 is
extended on a scale $\sim 2"$ (see below). GN20.2a is not spatially
resolved by these observations.  GN20.2b may be extended, although the
signal-to-noise of these data are not conclusive.

Figure 2 shows spectra of the three sources. For GN20, we integrate
over an area of $2"$. For the other two sources we only consider the
spectra at the peak of the velocity integrated CO image. Results for
Gaussian fits to the lines are given in Table 1. The lines are broad
for GN20 and GN20.2a ($\sim 800$ km s$^{-1}$), although not outside
the range of the zero intesity line widths seen in some nearby ULIRGs
(Downes \& Solomon 1998). However, we note athat, due to the bandwidth
limitation for the early EVLA, the very lowest frequency (highest
velocity) emission for GN20 may be truncated by these observations,
although the higher order CO transitions show that we expect little
emission beyond our lowest spectral channel (Daddi et
al. 2009a). GN20.2b shows a narrower profile at the peak
position. However, the detection in the velocity integrated image over
the full frequency range (Figure 1), suggests that the emission may be
extended both spatially and spectrally. A possibly broader line is
also suggested for the higher order CO emission, but again, the high
order CO detection is only marginal (Daddi et al 2009a).  We have
investigated this possibility with the current data, and conclude that
more sensitive observations are required to better characterize the CO
emission from GN20.2b, and we simply present the spectrum at the peak
position herein.

Figure 3 shows the velocity channel images for the GN20 CO 2-1
emission at 78 km s$^{-1}$ channel$^{-1}$ and $1.9"$ resolution.  The
emission clearly moves from the south to the north with increasing
frequency. An interesting feature is seen in channel 6, corresponding
to a frequency of 45.598 GHz (or $z =4.05588$).  The emission in
channel 6 appears to be substantially more extended than in other
channels. Inspecting our bandpass calibration, we find no
channel-dependent calibration errors that would lead to such a
difference in morphology for this single channel.  In channel 6 the
emission extends well to the north, with a total of extent of $\sim
4"$. Admittedly, the signal-to-noise is not high, but this extended
emission is suggestive of a gravitationally induced tidal feature.
Such tidal tails in nearby galaxies often occur over a narrow velocity
range (Hibbard \& Mihos 1995).

Figure 4a shows the CO 2-1 emission from GN20 integrated over the
velocity range shown in the channel maps of Figure 3. The emission
appears extended, in particular to the southwest.  A formal Gaussian
fit to the emission yields a total flux density of $0.69 \pm .07$ mJy,
a peak surface brightness of $0.49\pm 0.03$ Jy beam$^{-1}$, and a
deconvolved source size of $1.5" \times 0.9"$ with major axis position
angle PA = 69$^o$.  Figure 4b shows the iso-velocity contours
(first moment) of the CO 2-1 emission, ie. the weighted mean velocity
derived after first blanking the channel images below 3$\sigma$ per
channel. There is a velocity gradient north-south, with a magnitude of
$\pm 250$ km s$^{-1}$.  Given the low spatial resolution of these
images, this corresponds to a lower limit to the true projected
rotational velocity.

\section{Analysis}

\subsection{Masses}

We derive the mass in molecular gas (H$_2$) from the observed CO 2-1
luminosities using the standard relationships in Solomon \& Vanden
Bout (2005). We extrapolate from CO 2-1 to 1-0 luminosity assuming
constant brightness temperature, which is certainly correct to within
10\% for these galaxies (Carilli et al. 2010).  We adopt a CO
luminosity to H$_2$ conversion factor of $\alpha = 0.8$ M$_\odot$ K km
s$^{-1}$ pc$^2$, appropriate for nearby nuclear starburst galaxies
(Downes \& Solomon 1998).  The resulting values are listed in Table 1. 

Admittedly, the uncertainty on $\alpha$ is significant.
A minimum gas mass can be derived assuming optically thin
emission, and adopting a temperature and a CO to H$_2$ abundance
ratio.  Ivison et al. (2010) show that for reasonable assumptions
(Galactic CO abundance in molecular clouds, and temperatures ~ 40 K to
50 K), the lower limit to $\alpha \sim 0.65$ (see also Aalto et
al. 1995).

The gas mass in GN20 is $1.3\times 10^{11} \times (\alpha/0.8)$ M$_\odot$.
Daddi et al. (2009a) derive a stellar mass of $2.3\times 10^{11}$
M$_\odot$ from IR through optical SED fitting.

We can obtain a very rough estimate of the gravitational mass of GN20
from the CO velocity field, although there are substantial
uncertainties due to the low spatial resolution of these observations.
For the rotational velocity, we adopt the extremes of the velocity
channels with significant emission in Figure 3, or $390/sin(i)$ km
s$^{-1}$. For the radius, we adopt the value for the CO ring seen in
the high resolution 2-1 observations presented in Carilli et al.
(2010), $\sim 5$ kpc. The inclination angle remains uncertain.  The
gravitational mass inside this radius is $\sim 3.4\times 10^{11}
[sin(i)/sin(45^o)]^{-2}$ M$_\odot$. This naively derived dynamical
mass is comparable to the sum of the stellar and gas masses.

\subsection{Lyman-break galaxy molecular mass limits}

There is one Lyman-break galaxy within our field with a
spectroscopic redshift placing the CO 2-1 line within our band:
J123711.48+622155.8 at $z = 4.058$ (BD29079; Daddi et al. 2009a; Shim
et al. 2011).  We find no CO emission from this galaxy, to a $1\sigma$
limit of 0.17 mJy beam$^{-1}$ at 78 km s$^{-1}$ channel$^{-1}$, after
primary beam correction. Convolving to 312 km s$^{-1}$, we set a
3$\sigma$ limit to the CO 1-0 luminosity of $L'_{CO1-0} < 1.4\times
10^{10} \rm K~ km~ s^{-1}~ pc^2$, assuming constant brightness temperature
from CO 2-1 to 1-0. The implied molecular gas mass of this galaxy is
$< 1.1\times 10^{10} \times (\alpha/0.8)$ M$_\odot$. The stellar mass
of this galaxy is $2.6\times 10^{10}$ M$_\odot$ (Daddi et al. 2009a),
hence the gas-to-stellar mass ratio is $< 0.53 \times (\alpha/0.8)$.

Daddi et al. (2009a) estimate a star formation rate of 150 M$_\odot$
year$^{-1}$ for BD29079, or an IR luminosity of $1.5\times 10^{12}$
L$_\odot$. Daddi et al. (2010) have considered the 'star formation
law' in high $z$ galaxies, ie. the relationship between IR luminosity
and CO luminosity, and find for galaxies of this luminosity at $z \sim
2$: $L'_{CO1-0} \sim 0.02 L_{FIR}~\rm K~ km~ s^{-1}~ pc^2 $.  Adopting
this relation implies an expected CO luminosity of $L'_{CO1-0} \sim
3\times 10^{10}$ L$_\odot$. Hence, BD29079 is under-luminous in CO
emission relative to the standard star formation law, although within
the broad (factor three) scatter at any given luminosity.

There are 11 other LBGs within our field, and with photometric
redshifts that are consistent with the sampled range, although the
$z_{phot}$ error bars are such that the galaxies could be outside the
velocity range of the CO observations. We have looked for CO emission
at all these positions, and find no significant detections to similar
limits to those quoted above.

We emphasize that there is a large uncertainty in the derived H$_2$
mass for the LBGs, due to the unknown conversion factor, $\alpha$. It
is possible that metal poor star forming galaxies have a much larger
value of $\alpha$ than ULIRGs or even the Milky Way due to CO
dissociation by the more pervasive UV radiation field (Madden et
al. 1997; Papadopoulos \& Pelupessy 2010).

\section{Discussion}

We have detected CO 2-1 emission from three galaxies in the GN20
proto-cluster at $z = 4.05$ using the EVLA.  The molecular gas masses
range from $1.9\times 10^{10} \times (\alpha/0.8)$ M$_\odot$ to
$1.3\times 10^{11} \times (\alpha/0.8)$ M$_\odot$. Hence, we are
observing a group of molecular gas rich galaxies undergoing extreme
starburts within 1.6 Gyr of the Big Bang.

GN20 presents a particularly interesting case, given its very high
luminosity, and large spatial extent. We find that the sum of the
stellar and gas mass in GN20 is comparable to the dynamical
mass. While the assumptions involved in all three mass calculations
are uncertain, it appears that the baryons likely dominate the mass
content within 5 kpc radius of GN20.  Moreover, their is little room
for a CO luminosity to gas mass conversion factor substantially larger
than 0.8 M$_\odot$ K km s$^{-1}$ pc$^2$. High resolution observations
are required to confirm the rotational aspect of the emission, and to
obtain a more accurate derivation of the dynamical mass.

The morphology and velocity field for GN20 suggest a rotating disk,
with a possible tidal tail extending 10kpc to the north in a single 78
km s$^{-1}$ channel. We can speculate that GN20 is a gas rich disk
galaxy that has been gravitationally torqued by its neighboring
galaxies. This gravitational disturbance has not distrupted the
disk, but has greatly enhanced star formation in the disk, as
well as generating the extended, tidal gas distribution. 

Lastly, we have searched for CO emission from LBGs in the GN20
protocluster. We set an upper limit of $1.1\times 10^{10} \times
(\alpha/0.8)$ M$_\odot$ to the molecular gas mass in these galaxies,
including one with an accurate spectroscopic redshift. Our limit is
below the expected value based on the standard star formation 
law at this redshift, but within the broad scatter. 

We are obtaining high resolution CO imaging observations of the GN20
proto-cluster with the EVLA and the Plateau de Bure interferometer to
perform a more detailed dynamical analysis, and study the spatial
distribution of the CO excitation and gas-to-dust ratio of this
forming cluster of massive galaxies at $z = 4.05$.

\acknowledgments We thank the referee for useful comments. 
ED acknowledges the funding support of
ERC-StG-UPGAL-240039, and ANR-08-JCJC-0008. DR acknowledges support
from from NASA through Hubble Fellowship grant HST-HF-51235.01 awarded
by the Space Telescope Science Institute, which is operated by the
Association of Universities for Research in Astronomy, Inc., for NASA,
under contract NAS 5-26555.


\clearpage
\newpage

\begin{table}\label{}
\caption{Results from Gaussian fitting to the CO 2-1 line profiles}
\begin{tabular}[ht]{|c|c|c|c|c|c|c|c|}
\tableline
Source & Redshift & Peak & FWHM & I$_{CO}$ & Velocity$^a$ & L'CO & M(H$_2$) \\
~ & ~ & mJy & km s$^{-1}$ & Jy km s$^{-1}$ & km s$^{-1}$ & K km s$^{-1}$ pc$^2$ & M$_\odot$ \\
\tableline
GN20 & 4.0554 & 1.21$\pm$0.11  & 679$\pm$88 & 0.87$\pm$0.088 & -11$\pm$32 & 
$1.6\times 10^{11}$ & $1.3\times10^{11}$  \\
GN20.2a & 4.0508 & 0.54$\pm$0.065 & 723$\pm$110 & 0.41$\pm$0.050 & -283$\pm$45 &
$7.6\times 10^{10}$ & $6.1\times10^{10}$  \\
GN20.2b & 4.0567 & 0.42$\pm$0.083 & 290$\pm$70 & 0.13$\pm$0.025 & 67$\pm$29 &
$2.4\times 10^{10}$ & $1.9\times10^{10}$  \\
\tableline
\end{tabular}
$^a$Velocity relative to $z  = 4.0556$.
\end{table}

\clearpage
\newpage

\noindent  Aalto, S. et al. 1995, A\& A, 300, 369

\noindent Andreon, S. \& Huertas-Company, M. 2011, A\& A, 526, 11

\noindent  Aravena, M., Bertoldi, F., Carilli, C. et al. 2010, ApJL, 
708, L36

\noindent  Blain, A., Smail, I, Ivison, R., Kneib, J.-P., Frayer, D. 
2002, PhR 369 111

\noindent Capak, P. et al. 2011, Nature, 470, 233

\noindent  Carilli, C.L. \& Holdaway, M. 1999, Radio Science, 34, 817

\noindent  Casey, C.M., Chapman, S., Daddi, E. et al. 2009, MNRAS, in press

\noindent  Chapman, S., Blain, A., Ivison, R., Smail, I. 2003 Nature
  422 695

\noindent  Chapman, S., Blain, A., Ibata, R. et al. 2009, ApJ, 691, 560

\noindent  Collins, D., Stott, J.P., Hilton, M. et al. 2009, Nature,
458, 603

\noindent Cornwell, T., Braun, R., Briggs, D. 1999, in {\sl Synthesis
Imaging in Radio Astronomy II}, eds. G. B. Taylor, C. L. Carilli, and
R. A. Perley (ASP: San Francisco) 180, 151

\noindent  Daddi, E., Dickinson, M., Chary, R. et al 2005, ApJ 631 L13

\noindent  Daddi, E., Dannerbauer, H., Stern, D. et al. 
2009a, ApJ, 694, 1517

\noindent  Daddi, E., Dannerbauer, H., Krips, M. et al. 
2009b, ApJ, 695, L176

\noindent Daddi et al. 2010, ApJ, 714, L118

\noindent Dannerbauer, H., Walter, F., Morrison, G. 2008, ApJ, 673, L127

\noindent  Downes \& Solomon 1998 ApJ, 507, 615

\noindent  Doherty, M., Tanaka, M., de Breuck, C. et al.
2009, A\& A, 509, 83

\noindent Hibbard, J. \& Mihos, C. 1995, AJ, 110, 140

\noindent  Iono, D., Peck, A., Pope, A. et al. 2006, ApJ, 640, L1

\noindent  Ivison, R., Papadopoulos, P, Smail, I. et al. 2010,
  MNRAS, 412, 1913

\noindent  Kurk, J., Cimatti, A., Zamorani, G. et al. 2009,
A\& A, 504, 331

\noindent  Madden, S., Poglitsch, A., Geiss, N. et al. 1997, 483, 200

\noindent Momjian, E., Carilli, C. \& Petric, A. 2005, AJ, 129, 1809

\noindent Momjian, E. et al. 2010, AJ, 139, 1622

\noindent Moresco, M. et al. 2010, A\& A, 524, 67

\noindent Morrison, G., Owen, F., Dickinson, M., Ivison, R., Ibar,
E. 2010, ApJ, 188, S178

\noindent  Papadopoulos, P. \& Pelupessy, F. 2010, ApJ, 717, 1037

\noindent Perley, R.A. et al. 2011, ApJL, this volume

\noindent  Renzini, A. 2006, ARAA, 44, 141

\noindent  Riechers, D. et al. 2010, ApJL, 720, L131

\noindent  Solomon \& Vanden Bout 2005, ARAA, 43, 677

\noindent  Stevens, J, Ivison, R., Dunlop, J. 
et al. 2003 Nature, 425, 264

\noindent  Tacconi, L., Neri, R., Chapman, S. et al. 2006, ApJ, 640, 228

\noindent  Tacconi, L., Genzel, R., Smail, I., et al. 2008, ApJ, 680, 246

\noindent  Wardlow, J. et al. 2011, MNRAS, in press

\noindent  Williams, C.C., Giavaisco, M., Porciani, C.  et al. 2011,
ApJ, in press

\noindent  Younger, J., Fazio, G., Wilner, D. et al. 2008, 
ApJ, 688, 59

\clearpage
\newpage

\begin{figure}
\psfig{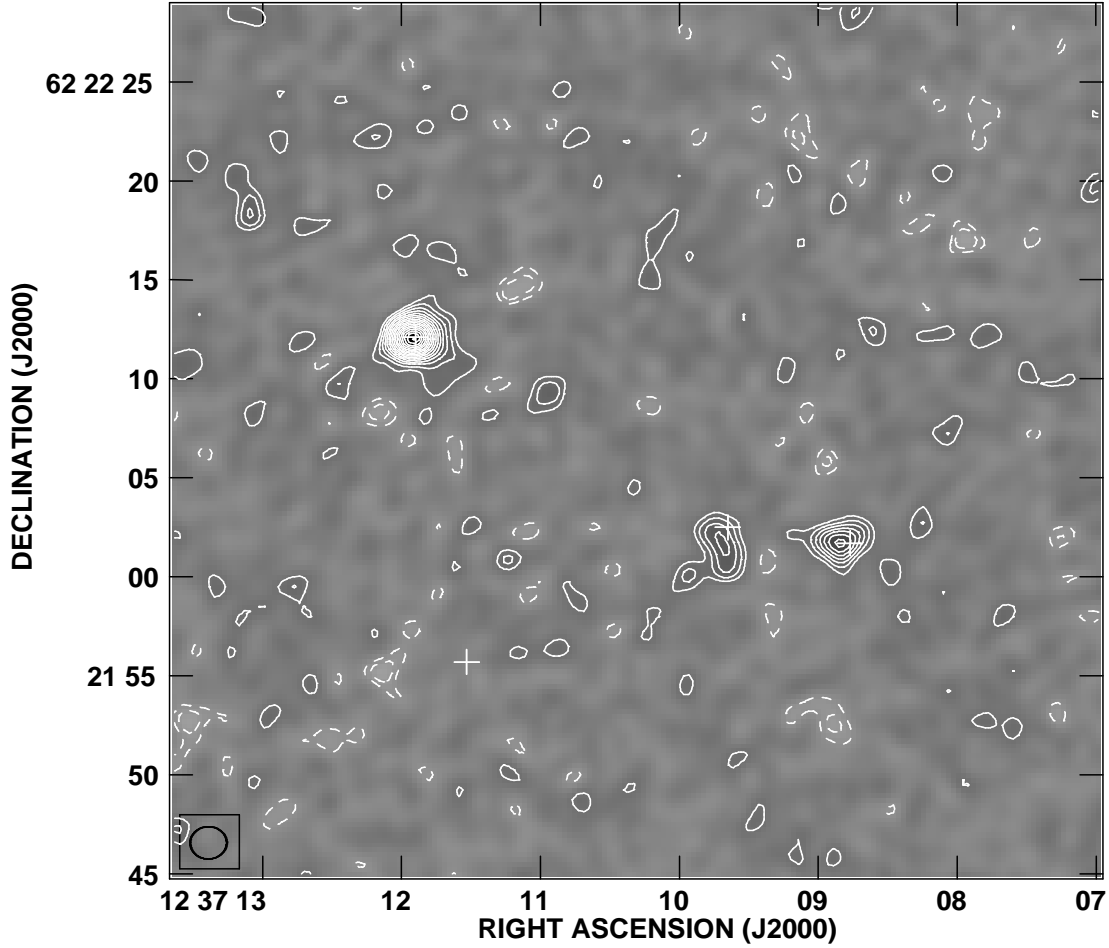}
\caption{
The CO 2-1 emission at $1.9"$ resolution integrated over the full
observed velocity range for the GN20 group of galaxies. In this case,
no primary correction is applied in order to ensure a flat noise
across the image to search for sources.  Crosses show the
1.4 GHz continuum positions of the three galaxies GN20, 20.2a,b (Morrison et
al. 2010), and the optical position of one LBG with a spectoscopic
redshift. The contour levels are linear, in steps of 1$\sigma$ (30 $\mu$Jy
beam$^{-1}$), starting at $\pm 2\sigma$.  Negative contours are
dashed.
}
\end{figure}

\begin{figure}
\psfig{file=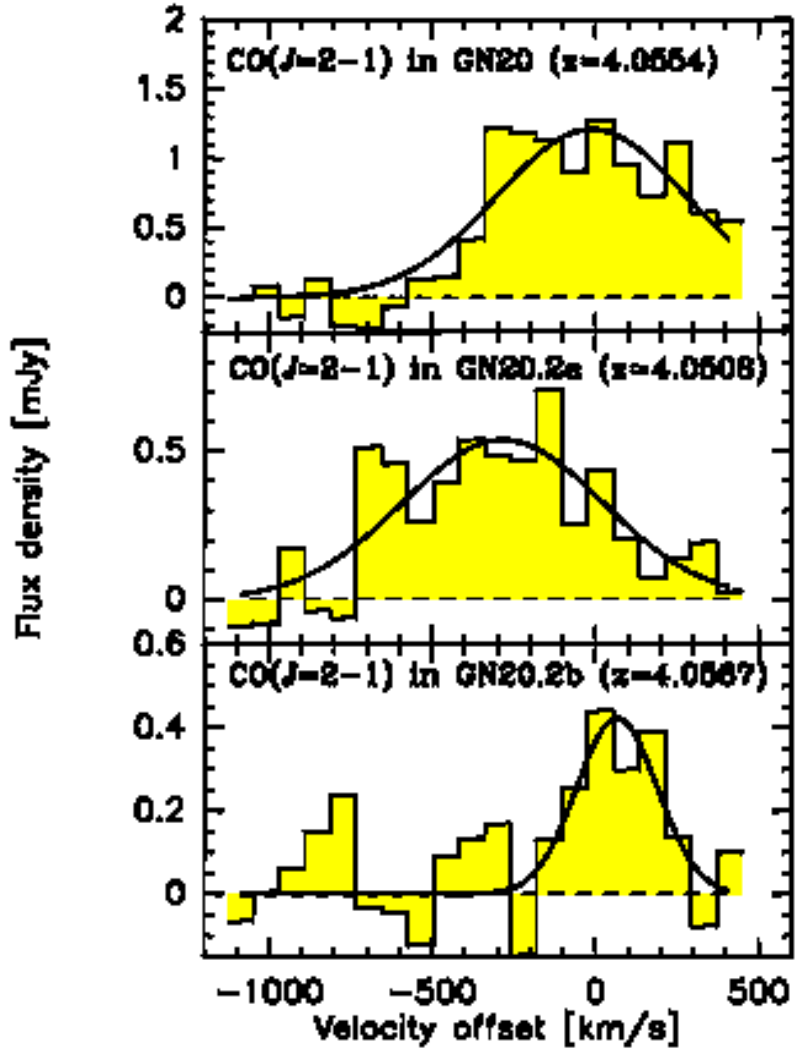,width=5in}
\caption{
CO 2-1 spectra for GN20, GN20.2a, and GN20.2b at 78 km s$^{-1}$ spectral
resolution.  Results for the Gaussian fitting are given in Table 1. 
}
\end{figure}

\begin{figure}
\psfig{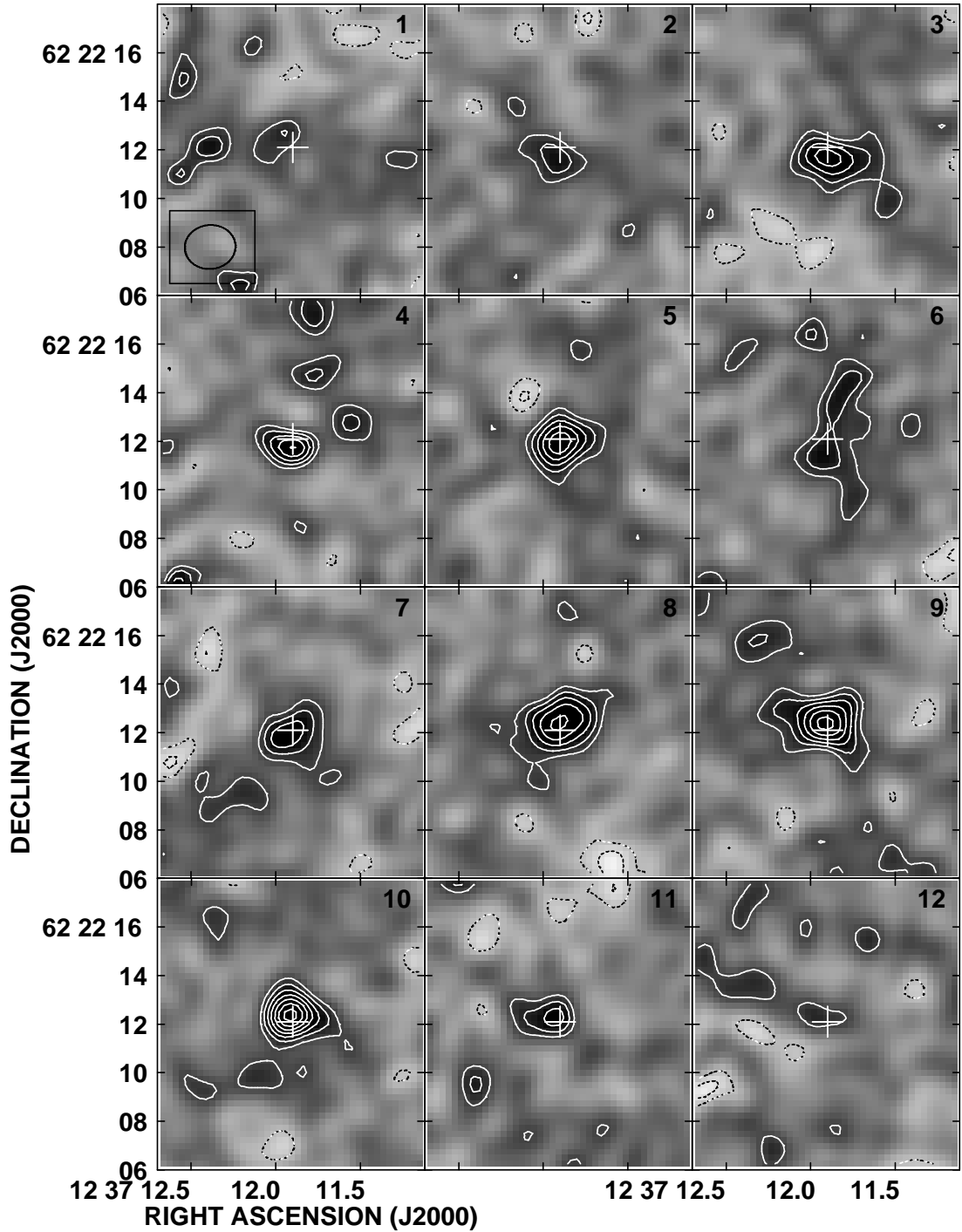}
\caption{
Channel images of the GN20 CO 2-1 spectrum at $1.9"$ resolution and
78 km s channel$^{-1}$. The contour
levels are linear, in steps of 1$\sigma$ (0.12 mJy beam$^{-1}$), starting
at $\pm 2\sigma$.  Negative contours are dashed.
}
\end{figure}

\begin{figure}
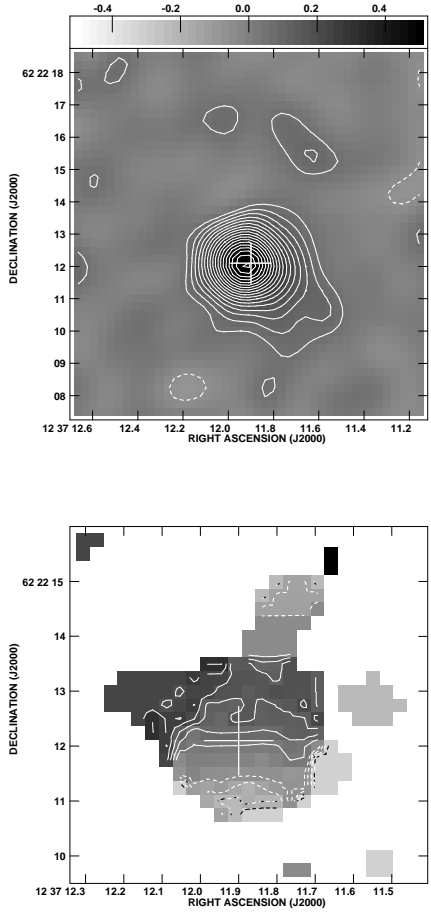

\psfig{file=Fig4a.ps,width=2.3in}
\psfig{file=Fig4b.ps,width=2.3in}
\caption{
Upper: Total intensity CO 2-1 image of GN20, summing all the line channels
(channels 1 to 12). The contour 
levels are linear, in steps of 1$\sigma$ (30$\mu$Jy beam$^{-1}$), starting
at $\pm 2\sigma$.  Negative contours are dashed.
Lower: Velocity contours of the CO 2-1 emission from GN20 (first moment image). 
Contours range from -250 km s$^{-1}$ to +250 km s$^{-1}$, in steps of 
50 km s$^{-1}$, with zero velocity corresponding to $z = 4.0556$.
Negative velocities are dashed. 
}
\end{figure}

\end{document}